\documentstyle[preprint,aps,epsfig]{revtex}
\draft
\tighten
\begin{document}

\title {Calculations of three-body observables in $^8$B breakup}
\author{J.\ A. Tostevin$^1$, F.\ M. Nunes$^{2,3}$, and I.\ J. Thompson$^1$}
\address{1) Department of Physics, School of Physics and Chemistry, University
of Surrey, Guildford, Surrey, GU2 7XH U.K.}
\address{2) Universidade Fernando Pessoa, Pra\c{c}a 9 de Abril, 4200 Porto,
Portugal}
\address{3) CENTRA, Instituto Superior T\'ecnico, Av.\ Rovisco Pais, 1096
Lisboa, Portugal}

\maketitle

\begin{abstract}
We discuss calculations of three-body observables for the breakup of $^8$B on a
$^{58}$Ni target at low energy using the coupled discretised continuum channels
approach. Calculations of both the angular distribution of the $^7$Be fragments
and their energy distributions are compared with those measured at several
laboratory angles. In these observables there is interference between the
breakup amplitudes from different spin-parity excitations of the projectile.
The resulting angle and the energy distributions reveal the importance of the
higher-order continuum state couplings for an understanding of the
measurements.
\end{abstract}
\date{\today}

\pacs{PACS: 24.10.Eq, 25.60.Gc, 25.70.De, 27.20.+n}

\section{Introduction}

Projectile breakup is an important reaction channel in the scattering of weakly
bound nuclei. An accurate treatment of breakup is therefore a major ingredient
in attempts to understand the properties of light exotic radioactive nuclei
from reaction studies. The number of published experimental breakup studies,
and also their accuracy, has increased rapidly. These include reactions in
which both nuclear and Coulomb breakup effects are expected to be significant,
e.g.
\cite{kaba,anne,blank,orr,riis,moto,naka1,bazin1,schwab,nd0,keely,kiku,bazin2,davids,naka2,iwasa,nd1}.
Until recently, the low intensities of available rare isotope beams has meant
that many of the experiments were either designed to measure inclusive cross
sections, with incomplete kinematics, or did not have adequate statistics to
allow the extraction of exclusive observables.  The cross sections extracted
from the measurements were often integrated over fragment energies or angles,
or both, and inevitably some details of the physical process were lost as a
result. This is no longer the situation. Secondary beam intensities are
becoming sufficiently high that coincidence experiments are now practical and,
in the future, data will more routinely require a fully three- or more-body
study, e.g. \cite{he6,b14}. The need for precise theoretical predictions of the
breakup of two-body projectiles, and of their three-body observables, is the
primary motivation for this work.

Theoretical reaction models, which treat breakup as an excitation of the
projectile to a two-body continuum state, most naturally express their results
as cross sections describing the center of mass (c.m.) and relative motions of
the dissociated system, using 2-body kinematics. It has therefore been common
for the experimental data to also be transformed to the c.m.\ frame, for ease
of comparison, e.g. the theoretical calculations of \cite{nunes1,esb2} and the
experimental data of \cite{nd1}. This process is ambiguous in the case of
inclusive data. Much more important is that the three-body cross sections are
explicitly coherent in contributions from different spin-parity excitations of
the projectile and so have the potential to offer a far greater insight into
the projectile structure and the reaction mechanism. An excellent example of
this is the interference observed \cite{davids} in the cross section of the
$^7$Be fragments, as a function of their component of momentum parallel to the
beam direction, following $^8$B breakup on a heavy target at 44 MeV/nucleon.

In this paper we present calculations which are performed using full three-body
kinematics. These calculations are carried out within the framework of the
coupled discretised continuum channels (CDCC) methodology,
e.g.\ \cite{PTP,PhysRep}, for breakup reactions of two-body projectiles. The
interference between different excitation channels is shown to be significant
for assessing the convergence of the calculations and those breakup excitations
which contribute. The methods presented are applied to the breakup of $^8$B on
a $^{58}$Ni target at $E_{lab}=25.8$ MeV, for which new measurements have been
reported \cite{nd1,nd2}.  We compare the results of the full CDCC analysis, and
also distorted waves Born approximation (DWBA) and truncated coupled channels
calculations, with these available data for the laboratory angle and energy
distributions of the $^7$Be fragments.  The calculations of Refs.\
\cite{nunes1,esb2} showed the importance of higher-order breakup couplings, the
couplings between continuum states, upon the $^8$B$^*$ center of mass cross
section angular distribution. We will show in this work that these higher-order
effects are manifest even more significantly in the energy distributions of the
$^7$Be fragments following breakup.

\section{Theoretical Considerations}

We consider the breakup reaction $p \rightarrow c+v$ in which the projectile
nucleus $p$ is a bound state of a core particle $c$, of spin $I$ and projection
$\mu$, and a valence particle $v$, of spin $s$ and projection $\sigma$. These
particles are, presently, assumed structureless and so their internal wave
functions are represented by the spinors ${\cal X}_I$ and ${\cal X}_s$. The
total angular momentum of the ground state of $p$ is $J_p$, with projection
$M$, the relative orbital angular momentum of the two constituents is $\ell_0$,
and their separation energy is ${\cal E}_0\ (>0)$. The incident wave number of
the projectile in the center of mass (c.m.) frame of the projectile and target
is $\vec{K}_0$ and the co-ordinate $z$--axis is chosen in the incident beam
direction. The target $t$ is assumed to have spin zero and no explicit target
excitation is included. Target excitation is therefore present only through the
complex effective interactions of $c$ and $v$ with the target. Our three-body
solution of the Schr\"{o}dinger equation calculates an approximate description
of the projection of the full many-body $p+t$ wave function onto the ground
states of the target, core and valence nuclei. This three-body wave function is
denoted $\Psi_{\vec{K}_0 M}(\vec{r}, \vec{R})$ where $\vec{R}$ is the position
of the c.m.\ of $p$ relative to the target and $\vec{r}$ is the position of $v$
relative to the core $c$. The particle masses are $m_p=m_c+m_v$ and $m_t$.

\subsection{Construction of continuum bin states}

In the coupled discretised continuum channels (CDCC) method \cite{PTP,PhysRep}
the breakup of $p$ is assumed to populate a finite set of selected $c+v$
excited configurations, with quantum numbers $J_p ',\ell, j$, where
$\vec{j}=\vec{\ell}+\vec{s}$ and $\vec{J}_p '=\vec{j}+\vec{I}$. Here, each of
these spin-parity excitations will be assumed diagonal in all of these angular
momentum labels. The excitations are also assumed to extend to some maximum
relative energy ${\cal E}_{max} (J_p ')$ or wave number $k_{max}$. This
momentum range is then divided into a number ${\cal N} (J_p ')$ of intervals or
bins, each with a width $\Delta k_i=[k_i- k_{i-1}]$. We label each such
momentum bin by $\alpha \equiv (i,J_p ',\ell,j,s,I)$.

In each of these relative motion bins a single representative wave function
is constructed from those $c+v$ scattering states $f_{\alpha }(k,r)$ internal
to the bin, with assumed angular momentum coupling
\begin{eqnarray}
\hat{\phi}_{\alpha}^{M'}(\vec{r})= \left[\left[Y_\ell(\hat{\vec{r}})\otimes
{\cal X}_s \right]_j \otimes {\cal X}_I \right]_{J_p 'M'} \, u_{\alpha }(r)/r
~~.
\end{eqnarray}
The radial functions $u_\alpha$ are square integrable and are a superposition
\begin{eqnarray}
u_{\alpha}(r) = \sqrt{\frac{2}{\pi N_\alpha }} \int_{k_{i-1}}
^{k_{i}} g_\alpha(k) f_{\alpha }(k,r)\, dk~~
\end{eqnarray}
of the scattering states, eigenstates of the $c+v$ internal Hamiltonian $H_p$,
with weight function $g_\alpha(k)$.  $N_\alpha = \int_{k_{i-1}} ^{k_{i}}
|g_\alpha (k)|^2 \, dk$ is a normalisation constant. The $f_{\alpha }$ are
defined here such that, for $r\rightarrow\infty$,
\begin{equation}
f_{\alpha}(k,r) \rightarrow \left[\cos\delta_{\alpha}(k)F_\ell(kr)+\sin\delta_{
\alpha}(k) G_\ell(kr)\right]~,
\end{equation}
where $k \in \alpha$ and $F_\ell$ and $G_\ell$ are the regular and irregular
partial wave Coulomb functions. So the $f_\alpha$ are real when using a real
$c+v$ two-body interaction. An optimal discretisation of the continuum requires
a consideration of the number, the boundaries $k_i$, the widths $\Delta k_i$
and the weights $g_\alpha$ in the bins, which may depend on the $J_p '$
configuration. Energy conservation relates the $c+v$ c.m.\ wave numbers
$K_\alpha$ and corresponding bin state energies $\hat{{\cal E}}_\alpha$, as
\begin{eqnarray}
\frac{\hbar^2 K_\alpha^2}{2\mu_{pt}}+\hat{{\cal E}}_\alpha = \frac{\hbar^2
K_0^2} {2\mu_{pt}}-{\cal E}_0~,
\end{eqnarray}
where we define each bin energy by $\hat{{\cal E}}_\alpha = \langle
\hat{\phi}_{\alpha}|H_p| \hat{\phi}_{\alpha}\rangle$ and where $\mu_{pt}$ is
the projectile-target reduced mass.

For non-$s$-wave bins typically one uses $g_\alpha(k)=1$ for a non-resonant
continuum in which case $N_i={\Delta k_i}$ and $\hat{{\cal E}}_i=
\hbar^2\hat{k}_i^2/(2\mu_{cv})$ with $\hat{k}_i^2= [k_i^3-k_{i-1}^3]/(3\Delta
k_i)$. For $s$-wave bins it is an advantage to use $g_\alpha(k)=k$. This
stabilises the extraction of the three-body transition amplitude at low
relative breakup energies, discussed later in Eq.\ (\ref{fff}). In this case
$N_i=\hat{k}_i^2 {\Delta k_i}$ and the bin energies are $\hat{{\cal E}}_i=
\hbar^2[k_i^5-k_{i-1}^5]/ (10 \mu_{cv} \Delta k_i \hat{k}_i^2)$.

\subsection{Coupled channels amplitudes}

These bin states $\hat{\phi}_{\alpha}$ provide an orthonormal relative motion
basis for the coupled channels solution of the three-body $c+v+t$ wave
function.  The bins and the coupling potentials $\langle \hat{\phi}_{\alpha}|
U(\vec{r},\vec{R})| \hat{\phi}_{\beta}\rangle$ are constructed, and the coupled
equations are solved, using the coupled channels code {\sc fresco}
\cite{fresco}. Here $U(\vec{r}, \vec{R})$ is the sum of the interactions of $c$
and $v$ with the target, which is expanded to a maximum specified multipole
order $q$.  The coupled equations solution generates the (two-body) scattering
amplitudes, summed over partial waves, for populating each bin state $J_p ',M'$
from initial state $J_p,M$, as a function of the angle $\theta_{K}$ of the
c.m.\ of the excited projectile in the c.m.\ frame
\begin{eqnarray}
\widehat{\cal
F}_{M'M}(\vec{K}_\alpha)&=&\frac{4\pi}{K_0}\,\sqrt{\frac{{K}_\alpha} {K_0}} \,
\sum_{LL'J} (L0J_p M|J M) \,(L'M-M'J_p 'M'|J M)\nonumber\\ &\times&
\exp(i[\sigma_L+\sigma_{L'}])\,\frac{1}{2i}\,\widehat{\cal S}^J_{LJ_p:  L'J_p
'}(K_\alpha)\,
Y_L^0(\widehat{\vec{K}}_0)\,Y_{L'}^{M-M'}(\widehat{\vec{K}}_\alpha)~~.\label{one}
\end{eqnarray}
Here $\sigma_L$ and $\sigma_{L'}$ are the Coulomb phases appropriate to the
initial and final state c.m.\ energies and the $\widehat{\cal S}_{L
J_p:L'J_p'}(K_\alpha)$ are the partial wave $S$--matrices for exciting bin
state $\alpha$ with c.m.\ wave number $K_\alpha$.  When calculated using {\sc
fresco} \cite{fresco}, these amplitudes are expressed in a coordinate system
with $x$--axis in the plane of $\vec{K}_0$ and $\vec{K}_\alpha$, such that the
azimuthal angle $\phi_{{K}_\alpha}$ of $\vec{K}_\alpha$ is zero. When
discussing three-body observables, it is more convenient to define the
coordinate system with respect to the fixed positions of the detectors in the
laboratory. For such a general $x$--coordinate axis the coupled channels
amplitudes must subsequently be multiplied by $\exp(i[M-M'] \phi_{ {K}})$, with
$\phi_K$ referred to the chosen $x$-axis.

For use in the following, the two-body inelastic amplitudes of Eq.\ (\ref{one})
are re-normalised to that of the $T$-matrix by removal of their two-body phase
space factors, so that
\begin{eqnarray}
\widehat{\cal T}_{M'M}^\alpha(\vec{K}_\alpha)=-\frac{2\pi\hbar^2}{\mu_{pt}}\,
\sqrt{\frac{K_0}{K_\alpha}} \, \widehat{\cal F}_{M'M}(\vec{K}_\alpha)~~.
\label{renorm}
\end{eqnarray}
Throughout, we adopt scattering state and $T$-matrix normalisations such that,
asymptotically, the plane wave states $\exp(i\vec{k}\cdot\vec{r})$ that enter
are multiplied by unity.

It follows that the inelastic differential cross section angular distribution,
in the center of mass frame, for excitation of a given bin state is
\begin{eqnarray}
\frac{d\sigma (\alpha)}{d\Omega_{K}} &=& \frac{1}{2J_p+1}\,\left[
\frac{\mu_{pt}}{2\pi\hbar^2} \right]^2 \,\frac{K_\alpha}{K_0} \sum_{MM'} \left|
\widehat{\cal T}_{M'M}^\alpha (\vec{K}_\alpha) \right|^2 \nonumber \\ &=&
\frac{1}{2J_p+1}\, \sum_{MM'} \left| \widehat{\cal F}_{M'M}(\vec{K}_\alpha)
\right|^2 ~~.\label{doub}
\end{eqnarray}

\subsection{Three-body breakup amplitudes}

Less obvious is the relationship of the CDCC two-body inelastic amplitudes
$\widehat{\cal T}_{M'M}^\alpha (\vec{K}_\alpha)$ to the breakup transition
amplitudes $T_{\mu\sigma:M} (\vec{k},\vec{K})$ from an initial state $J_p,M$ to
a general physical three-body final state of the constituents \cite{PTP,book}.
This is needed to make predictions for experiments with general detection
geometries, since each detector configuration and detected fragment energy
involves a distinct final state c.m.\ wave vector $\vec{K}$, breakup energy
${\cal E}_k$, and relative motion wave vector $\vec{k}$.

To clarify this connection we make the CDCC approximation to the exact (prior
form) breakup transition amplitude, by replacing the exact $c+v+t$ three-body
wave function, $\Psi_{\vec{K}_0 M}(\vec{r},\vec{R})$, by its CDCC approximation
$\Psi^{CD}$, as
\begin{eqnarray}
T_{\mu\sigma:M}(\vec{k},\vec{K})=\langle \phi_{\vec{k}\mu\sigma}^{(-)}(\vec{r})
\,e^{i\vec{K} \cdot\vec{R}}|U(\vec{r},\vec{R})| \Psi_{\vec{K}_0
M}^{CD}(\vec{r},\vec{R}) \rangle~~.
\end{eqnarray}
Here $\phi_{\vec{k}\mu\sigma}$ is the $c+v$ final state. Upon inserting the set
of all included bin-states, which are assumed complete within the model space
used, then
\begin{eqnarray}
T_{\mu \sigma:M}(\vec{k},\vec{K})=\sum_{\alpha,M'} \langle
\phi_{\vec{k}\mu\sigma}^{(-)}|\hat{\phi}^{M'}_{\alpha} \rangle \langle
\hat{\phi}^{M'}_{\alpha}\,e^{i\vec{K} \cdot\vec{R}}|U(\vec{r},\vec{R})|
\Psi_{\vec{K}_0 M}^{CD}(\vec{r},\vec{R})\rangle~,\label{comp}
\end{eqnarray}
where the sum is over all bins $\alpha$ which contain wave number $k$.  We
should now recognise that the matrix elements $\widehat{\cal T}_{M'M}^\alpha
(\vec{K}_\alpha)$ of Eq.\ (\ref{renorm}), obtained from the coupled channels
solution, are precisely the transition matrix elements appearing in
Eq.\ (\ref{comp}), i.e.
\begin{eqnarray}
\widehat{\cal T}_{M'M}^\alpha(\vec{K}_\alpha)= \langle \hat{ \phi}^{ M'}_{
\alpha}\,e^{i\vec{K}_\alpha \cdot\vec{R}} |U(\vec{r},\vec{R})| \Psi_{\vec{K}_0
M}^{CD}(\vec{r},\vec{R})\rangle,
\end{eqnarray}
but calculated on the grid of $\theta_\alpha$ and $K_\alpha$ values. For the first
term in Eq.\ (\ref{comp}) one obtains
\begin{eqnarray}
\langle \phi_{\vec{k}\mu\sigma}^{(-)}|\hat{\phi}^{M'}_{\alpha}
\rangle=\frac{(2\pi)^{3/2}}{k\sqrt{N_\alpha}}\sum_{\nu}\,(-i)^\ell\,
(\ell\nu s\sigma |j m)\,(jm I\mu |J_p' M')\, \exp[i
\bar{\delta}_{\alpha}(k)]\, g_\alpha (k) Y_\ell^\nu (\hat{\vec{k}})~,
\end{eqnarray}
where $\bar{\delta}_{\alpha}(k)={\delta}_{\alpha}(k)+ \sigma_\alpha (k)$ is the
sum of the nuclear and Coulomb phase shifts for $c+v$ scattering at relative
wave number $k$. It follows that the three-body breakup $T$--matrix can be
written
\begin{eqnarray}
T_{\mu\sigma:M}(\vec{k},\vec{K})&=&\frac{(2\pi)^{3/2}}{k}\sum_{\alpha\,
\nu} \,(-i)^\ell\, (\ell\nu s\sigma |j m)\,(jm I\mu |J_p 'M')\, \exp[i
\bar{\delta}_{\alpha}(k)] \nonumber\\ &\times& Y_\ell^\nu
(\hat{\vec{k}})\, g_\alpha (k) \, {T}_{M'M}(\alpha,\vec{K})~.\label{fff}
\end{eqnarray}
Here the ${T}_{M'M}(\alpha,\vec{K})$ will be interpolated from the matrices
$\widehat{\cal T}_{M'M}^\alpha(\vec{K}_\alpha)$, available on the calculated
$K_\alpha$ and $\theta_{K_\alpha}$ grid. Specifically, for each value of
$\vec{K}$, we evaluate
\begin{eqnarray}
{T}_{M'M}(\alpha,\vec{K})= \exp(i[M-M']\phi_{K})\,\left[\widehat{\cal T}_{M'M}
^\alpha (\vec{K})/ \sqrt{N_\alpha}\,\,\right]~ \label{ggg}
\end{eqnarray}
where the value of the bracketed term on the right hand side is interpolated
from the coupled channels solution.

In practice this interpolation is carried out as a function of the deviation of
$K$ from the threshold center of mass wave number. For non-$s$-wave breakup,
the amplitudes are constrained to vanish at the breakup threshold $K_{thr}$,
i.e.
\begin{eqnarray}
\widehat{\cal T}_{M'M}^{\ell\neq 0}(\vec{K}_{thr})=0~,~~~~~ \frac{\hbar^2
K_{thr}^2} {2\mu_{pt}}= \frac{\hbar^2 K_0^2}{2\mu_{pt}}-{\cal E}_0~~.
\end{eqnarray}
We note that in Eqs.\ (\ref{fff}) and (\ref{ggg}) the functional dependence of
the $T$-matrix on the angles of $\vec{k}$, the phase shifts $\bar{ \delta}_{
\alpha}(k)$, and the azimuthal angle $\phi_K$ are all treated exactly. The grid
of $\theta_K$ values can also be very fine without computing cost. The most
important requirement is therefore that the number of bin states used to
describe each $[0\rightarrow k_{max}]$ $J_p '$ excitation must be sufficient to
allow an accurate interpolation of the amplitudes in the value of $\Delta
K=|K-K_{thr}|$, or alternatively in $k$.

\subsection{Three-body observables}

The three-body amplitudes of Eq.\ (\ref{fff}) are used to compute the triple
differential cross sections for breakup. If the energy of the core particle is
measured then
\begin{equation}
\frac{d^3\sigma}{d\Omega_c d\Omega_v dE_c}=\frac{2\pi\mu_{pt}}{\hbar^2 K_0}
\,\frac{1}{(2J_p+1)}\,\sum_{\mu\sigma M}\,\left|T_{\mu\sigma:M}(\vec{k},\vec{K})
\right|^2 \,\rho(E_c,\Omega_c,\Omega_v) \label{trip}~~.
\label{eq:sigma3}
\end{equation}
With our $T$-matrix normalisations, and non-relativistic kinematics, the
necessary three-body phase space factor $\rho(E_c,\Omega_c,\Omega_v)$, the
density of states per unit core particle energy interval for detection at solid
angles $\Omega_v$ and $\Omega_c$, is \cite{fuchs}
\begin{equation}
\rho(E_c,\Omega_c,\Omega_v)=\frac{m_c m_v\hbar k_c\hbar
k_v}{(2\pi\hbar)^6}\left[\frac{m_t} {m_v+m_t+m_v(\vec{k}_c-\vec{K}_{tot})\cdot
\vec{k}_v/k_v^2}\right]~.\label{phase}
\end{equation}
Here $\hbar\vec{k}_c$ and $\hbar\vec{k}_v$ are the core and valence particle
momenta in the final state and $\hbar\vec{K}_{tot}$ the total momentum of the
system, all evaluated in the frame, c.m.\ or laboratory, of interest. The
association with the appropriate $T$-matrix elements in Eq.\ (\ref{trip}) is
made through
\begin{equation}
\vec{K}=\vec{k}_c+\vec{k}_v-\frac{m_p}{m_p+m_t}\vec{K}_{tot}~,
\qquad \vec{k}=\frac{m_c}{m_p}\vec{k}_v-\frac{m_v}{m_p}\vec{k}_c~.
\end{equation}

As the data under discussion here are inclusive with respect to the valence
particle (proton) angles, the calculated triple differential cross sections
must be integrated numerically over $\Omega_v$. The presented observables are
also integrated and averaged over the solid angles $\Delta \Omega_c$ subtended
by the core particle detectors, with the stated detector efficiency profiles
$\varepsilon(\Omega_c)$ \cite{nd1}, i.e.
\begin{equation}
\left\langle \frac{d^2\sigma}{d\Omega_c dE_c}\right\rangle= \frac{1}{\Delta
\Omega_c} \int_{\Delta\Omega_c} d\Omega_c \, \left\{\varepsilon (\Omega_c)
\, \int d\Omega_v \frac{d^3 \sigma}{d\Omega_c d\Omega_v dE_c}\right\}~.
\end{equation}
It is most convenient to choose the $x-z$ plane to be that defined by the beam
and the core particle detector.

\section{Application to sub-Coulomb breakup}

The method detailed above is applied to the breakup of $^8$B on $^{58}$Ni at
energy $E_{lab} = 25.8$ MeV, for which new data are available \cite{nd1,nd2}.
A first experiment was performed in 1996 at the Nuclear Structure Laboratory of
the University of Notre Dame (ND) \cite{nd0}, one motivation being to clarify
the importance of the E2 contribution to the Coulomb dissociation process, an
issue which is still not completely resolved \cite{kiku}. In that first
experiment, the measured $^7$Be fragments were detected at only one laboratory
angle ($\approx 40^\circ$), assumed to be free from the influence of strong
interaction contributions. However, as a result of theoretical predictions
\cite{nunes2,dasso} of a strong nuclear peak beyond $40^\circ$, and claims also
of Coulomb-nuclear interference at around $40^\circ$, a more complete
experiment was recently carried out using the now upgraded ND facility.
Measurements were obtained of an angular distribution of the $^7$Be fragments
\cite{nd1} and also of their energy distributions \cite{nd2} for the range of
measured laboratory angles.  Although the removed proton is not observed, since
the heavy fragment energies are identified, the presented $^7$Be fragment
distributions are known to contain no contribution from proton transfer
reactions to bound states of $^{59}$Cu. There may nevertheless be contributions
from knockout or stripping processes in which the proton excites the target and
is absorbed. Such contributions are not calculated in this work.  Proton
transfer reactions to near-threshold (unbound) states of $^{59}$Cu, if present,
could also contribute.  We comment briefly below on the latter.

\subsection{The CDCC model space}

Model space parameters common to all the CDCC calculations are as follows.
Partial waves up to $L_{max}=1000$ and radii $R$ up to  $R_{coup}=500$ fm  were
used for the computation of the projectile-target relative motion wave
functions. The continuum bins were calculated using radii $r \leq r_{bin}=60$
fm. The $^7$Be intrinsic spin was neglected, the core being assumed to behave
as a spectator.  Thus we set $I=0$. The proton spin, $s=1/2$, was included and
hence $J_p '= j$.

In the final calculations presented all $J_p '$ states consistent with relative
orbital angular momenta $\ell\leq 3$, i.e. $J_p '$ up to $f_{7/2}$, were
included. We show that the effects of the $g$-wave continuum are small. The bin
state discretisation was carried out up to maximum relative energy ${\cal
E}_{max}=10$ MeV for each state. The number of bins in the $s_{1/2} $-continuum
was 32. For each of the other $J_p '$, 16 bins were used. These had equally
spaced $k_i$ from $k=0$ to $k_{max}$.  In the case of the DWBA calculations
shown the model space is the same, however, the bin states are coupled to the
ground state in first-order only. Calculations using potential multipoles
$q\leq 4$ in constructing the coupling potentials will be shown but the final
calculations require $q\leq 3$.

For the $^7$Be-$^{58}$Ni system, the interaction of Moroz {\em et al.}
\cite{moroz} was used, as in the earlier analysis of Ref.\ \cite{nunes1}. The
proton-$^7$Be binding potential was taken from Esbensen and Bertsch (EB)
\cite{esb}. The model of Kim {\em et al.} \cite{kim} is also considered. The
potential used to construct the bin states was the same (real) potential as was
used to bind the $^8$B ground state, assumed a pure $p_{3/2}$ proton
single-particle state. The proton-$^{58}$Ni potential is first taken from the
global parameterization of Becchetti and Greenlees (BG) \cite{becc}, but is
also discussed below.

\subsection{Results of calculations}

It is important to note from the outset that the total breakup cross section
angular distribution of the c.m.\ of the excited projectile, the sum of the
two-body inelastic differential cross sections of Eq.\ (\ref{doub}), is
incoherent in the different bin components. This is not the case for the
three-body amplitude of Eq.\ (\ref{fff}) and the triple differential cross
sections, Eq.\ (\ref{trip}). The practical convergence of the calculation,
i.e.  the dependence of the observables on the assumed model space, is
therefore much more subtle in this case.

The three-body calculations are found to require a more extended set of bins,
excitation energies, and potential multipoles.  Whereas the use of energy bins
up to only 3 MeV of relative energy, and multipoles $q\leq 2$, e.g.\ in
Ref.\ \cite{nunes1}, gives stable (converged) c.m.\ differential cross
sections, in the sense of Eq.\ (\ref{doub}), this is not the case for the
calculations of the triple differential cross sections and the energy and angle
integrated distributions.  We need the enlarged coupled channels model space,
as detailed above, with bins extending beyond ${\cal E}_{max}=8$ MeV to obtain
a converged result for these three-body observables. Furthermore, even when the
extended range of continuum energies is included, the bin discretisation may
itself not be fine enough so that the basis of bin states is sufficiently
complete. We have therefore verified the stability of our results, with regard
to the bin size, by doubling the number of bins and confirming that the same
results are produced.

\subsubsection{Angular distributions}

\begin{figure}[h]
\epsfig{figure=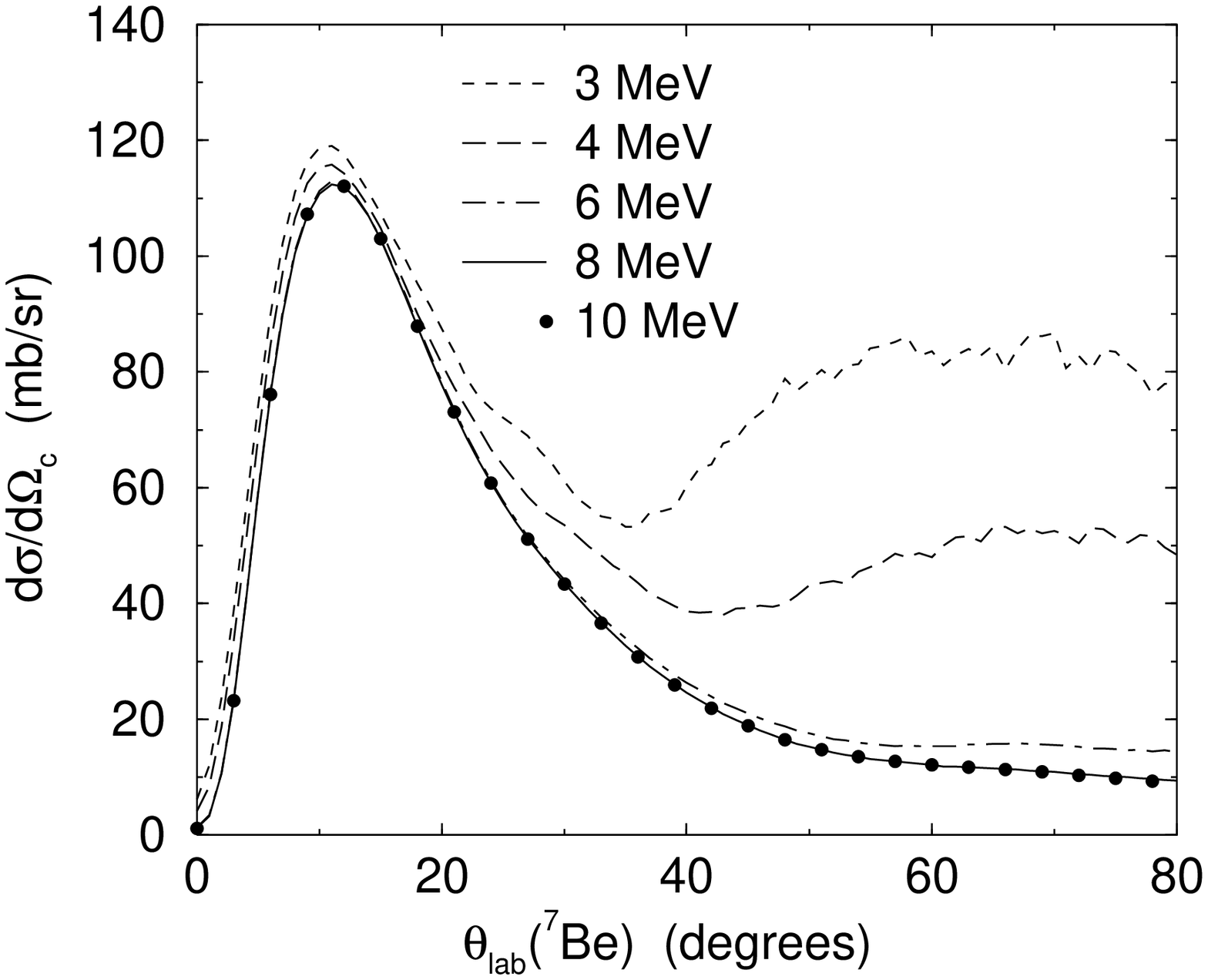,width=0.60\textwidth}
\medskip
\caption{Convergence of the calculated laboratory frame $^7$Be cross section
angular distribution following the breakup of $^8$B on $^{58}$Ni at 25.8 MeV as
a function of the maximum proton-$^7$Be relative energy included in the
calculation.} \label{fig:conv}
\end{figure}

The convergence of the three-body calculations with ${\cal E}_{max}$ is clearly
illustrated in Fig.\ \ref{fig:conv}. Here we show the $^7$Be laboratory
differential cross section angular distributions from calculations that include
continuum bins up to ${\cal E}_{max}=3,4,6,8$, and 10 MeV. The calculations for
this convergence test use multipoles $q\leq 2$ and $\ell \leq 3$. The
calculations use the BG proton-target potential and the EB proton-$^7$Be
potential.  For the larger ${\cal E}_{max}$ the bins have been constructed so
as not to alter their low energy discretisation. The calculation of the
three-body cross sections thus provides a different interpretation of the
reaction mechanism, and evidence for significantly higher energy excitations
than would be deduced from the earlier calculations and their comparison with
the $^8$B$^*$ c.m.\ cross section.  We will show that these high relative
motion excitations are reflected in the calculated breakup energy distributions
for $^7$Be and the proton.

\begin{figure}[h]
\epsfig{figure=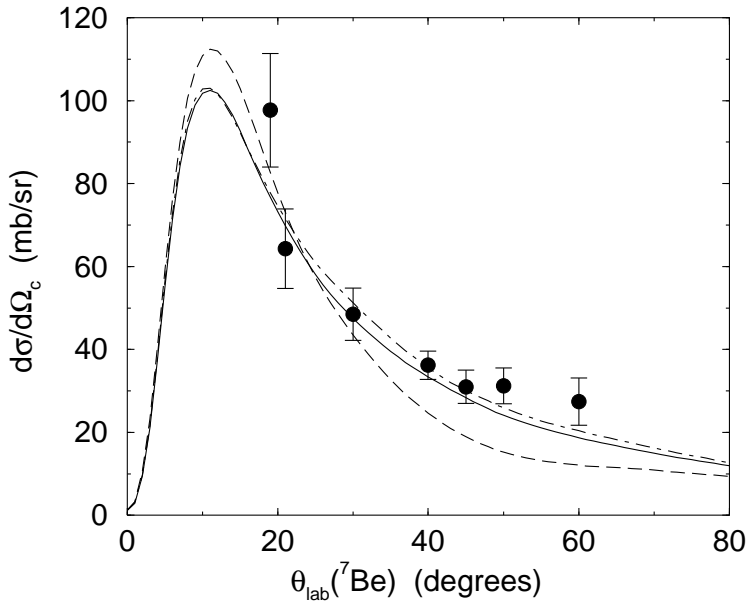,width=0.60\textwidth}
\medskip
\caption{The calculated laboratory frame $^7$Be cross section angular
distribution following the breakup of $^8$B on $^{58}$Ni at 25.8 MeV.  The
long-dashed curve is the ${\cal E}_{max}=10$ MeV, $\ell\leq 3$, $q\leq 2$,
calculation from Fig.\ \protect\ref{fig:conv}. The solid curve includes $q=3$
multipole terms while the dot-dashed curve includes both $q=4$ and $\ell=4$
effects.  } \label{fig:imag1}
\end{figure}

\begin{figure}[h]
\epsfig{figure=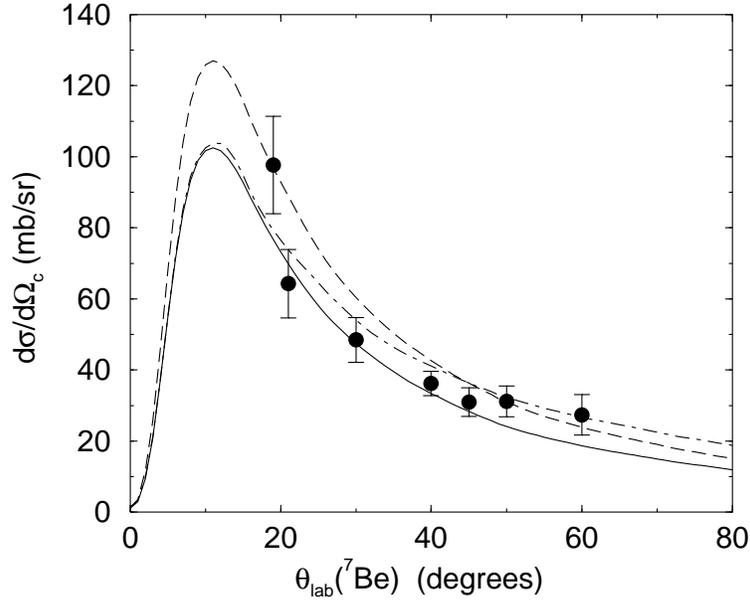,width=0.60\textwidth}
\medskip
\caption{The calculated laboratory frame $^7$Be cross section angular
distribution following the breakup of $^8$B on $^{58}$Ni at 25.8 MeV from the
EB (solid) and Kim (dashed) models for the  proton-$^7$Be interaction and the
BG proton-target interaction. The dot-dashed curve uses the EB proton-$^7$Be
interaction and the VG proton-target interaction.  The experimental data are
from Ref.\ \protect\cite{nd1}.}\label{fig:ang}
\end{figure}

Figs.\ \ref{fig:imag1} and \ref{fig:ang} present the calculated $^7$Be
laboratory differential cross section angular distribution, integrated over
energy and proton angles and averaged over the core detector solid angles, and
compare this with the data \cite{nd2}. The $^7$Be detectors were circular,
subtending a solid angle $\Delta\Omega_c$ comprising a circle of radius
$6^\circ$ about the nominal laboratory angle $\theta_{lab}$. They have a stated
Gaussian efficiency profile $\varepsilon (\theta)$ with full width half maximum
of 10.9$^\circ$ \cite{nd1}.  Here $\theta$ is measured from the nominal
$\theta_{lab}$ setting.

The convergence of the calculations with multipole order, and also with the
included continuum partial waves, is shown in Fig.\ \ref{fig:imag1}. Here the
long dashed curve is the result shown in Fig.\ \ref{fig:conv}, converged with
respect to excitation energy, with $q\leq 2$ and $\ell \leq 3$. The solid curve
includes also the effects of the $q=3$ multipole couplings for $\ell \leq 3$.
The dash-dot curve is a calculation where $q=4$ multipole couplings and the
$\ell=4$ breakup partial wave are included. The additional effects are small
and the remaining calculations therefore use the truncated model space with
$q\leq 3$ and $\ell \leq 3$.

The solid curve in Fig.\ \ref{fig:ang} uses the BG proton-target potential and
the EB proton-$^7$Be potential.  In Ref.\ \cite{nunes2} it was shown that
different $^7$Be-$^{58}$Ni potential models give essentially the same shape for
the $^8$B$^*$ c.m.\ angular distribution, while the cross section normalisation
depends on the size of the $^8$B g.s. wave function. The long-dashed curve in
Fig.\ \ref{fig:ang} shows the results of using the proton-$^7$Be interaction of
Kim {\em et al.} \cite{kim}. Consistent with earlier work, the cross section is
enhanced due to the larger predicted $^8$B root mean squared (r.m.s.) radius in
this model.

The Becchetti-Greenlees \cite{becc} proton-$^{58}$Ni potential, used above and
previously, has surface imaginary strength and geometry parameters $W$=12 MeV,
$r_W$=1.32 fm, and $a_W$=0.534 fm when computed at 3 MeV proton energy.
Experience tells us \cite{PP} that the BG parameters give reasonable fits to
data only down to approximately 10 MeV.  An alternative global
parameterization, tailored for use below 20 MeV, has a similar imaginary
strength but somewhat smaller radius and diffuseness parameters $r_W$=1.25 fm,
and $a_W$=0.47 fm \cite{perey} and leads to very similar results. There are,
however, also potential parameters fitted to elastic scattering data at 5.45
MeV \cite{van61,PP}. This analysis uses a Gaussian surface term and obtains a
much reduced absorptive strength, $W$=3.5 MeV, $r_W$=1.23 fm, and $a_W$=1.2
fm.  We will refer to this as the VG potential. There is therefore some
uncertainty in this potential input. The dot-dashed curve in Fig.\ \ref{fig:ang}
shows the calculated $^7$Be angular distribution from the VG potential.  The
cross section is changed only slightly at smaller angles.  At the larger
angles the calculated cross section is enhanced and is consistent with the
experimental angular distribution data.

Our calculations show that the $^8$B structure (size) and proton-target
potential uncertainties affect the calculations in characteristically different
ways. The former produces an overall scaling while the latter produces,
principally, a large angle enhancement. The data, currently, do not allow these
effects to be discriminated further.  In the final event, the overall agreement
between the calculations and the data in Fig.\ \ref{fig:ang} is qualitatively
similar to the comparisons made in Ref.\ \cite{nd1}. There the calculated
$^8$B$^*$ c.m.\ cross sections \cite{nunes1,esb2} are compared with an
approximate transformation of the measured $^7$Be data of Fig.\ \ref{fig:ang}
to the c.m.\ frame. Such approximate comparisons, however, are not necessary.

We observe that the results of our calculations are qualitatively quite
different to those presented in Ref.\ \cite{shyam}, where an isotropic
approximation was assumed in calculating the $^7$Be fragment laboratory cross
sections. Those calculations show a radical change of shape of the angular
distribution at forward angles which is not present in the calculations of
Figs.\ \ref{fig:conv}, \ref{fig:imag1}, and \ref{fig:ang} in which the angular
dependences are treated exactly.

\subsubsection{Energy distributions}

\begin{figure}[h]
\centerline{
\hspace{-0.5in}
        \parbox[t]{0.4\textwidth}{
\epsfig{figure=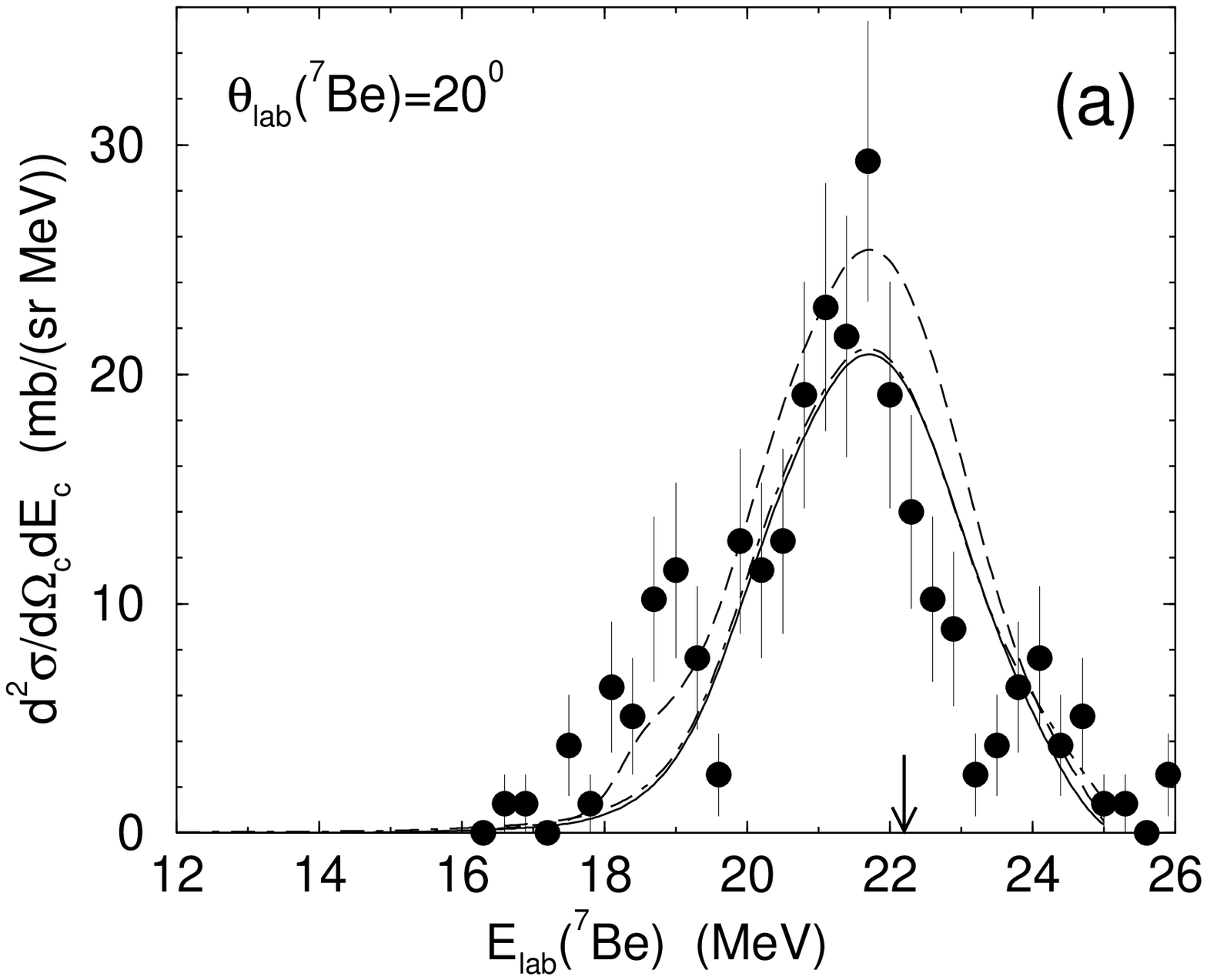,width=0.45\textwidth}}
\hspace{0.4in}
        \parbox[t]{0.4\textwidth}{
\epsfig{figure=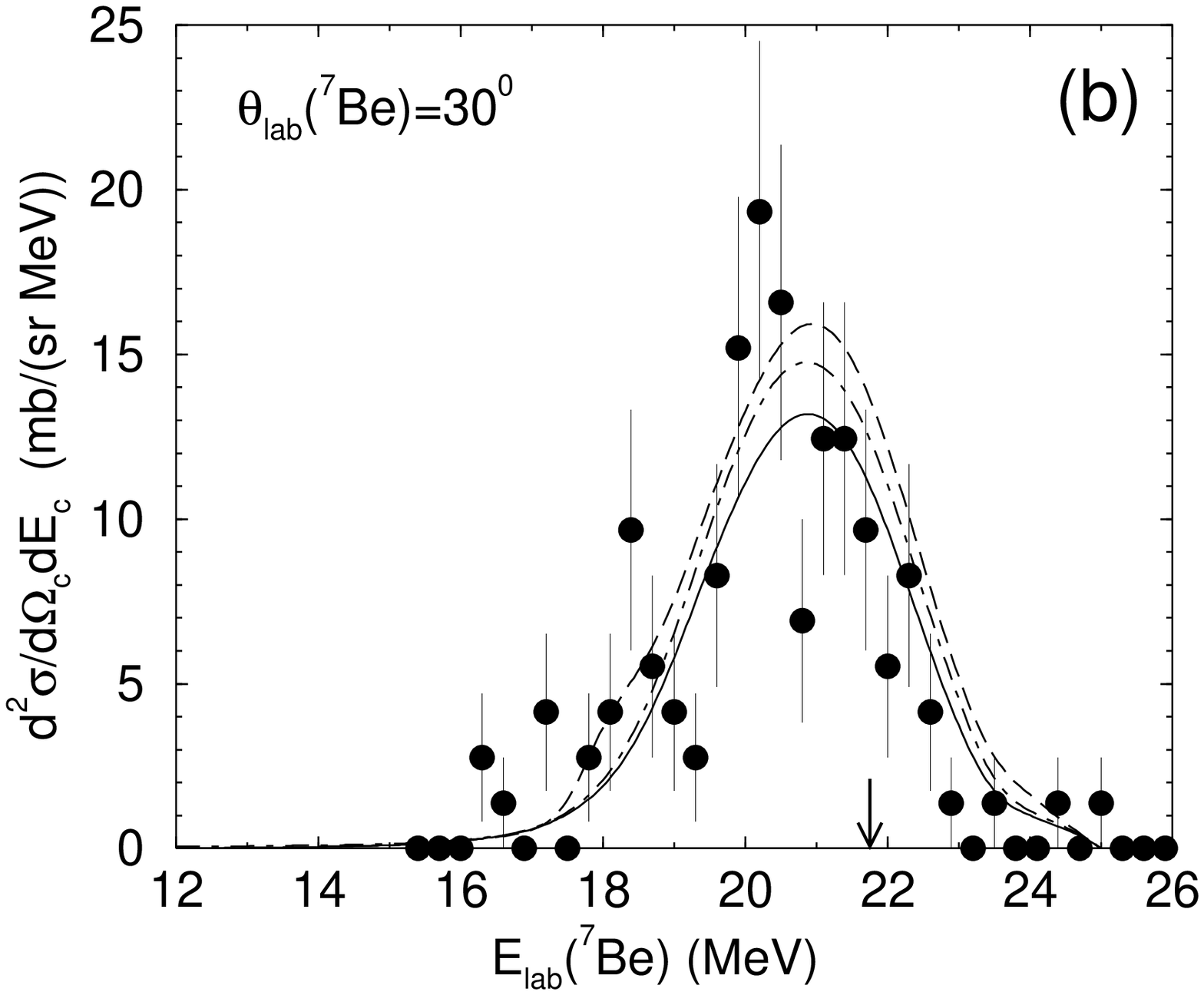,width=0.45\textwidth}}
}
\centerline{
\hspace{-0.5in}
        \parbox[t]{0.4\textwidth}{
\epsfig{figure=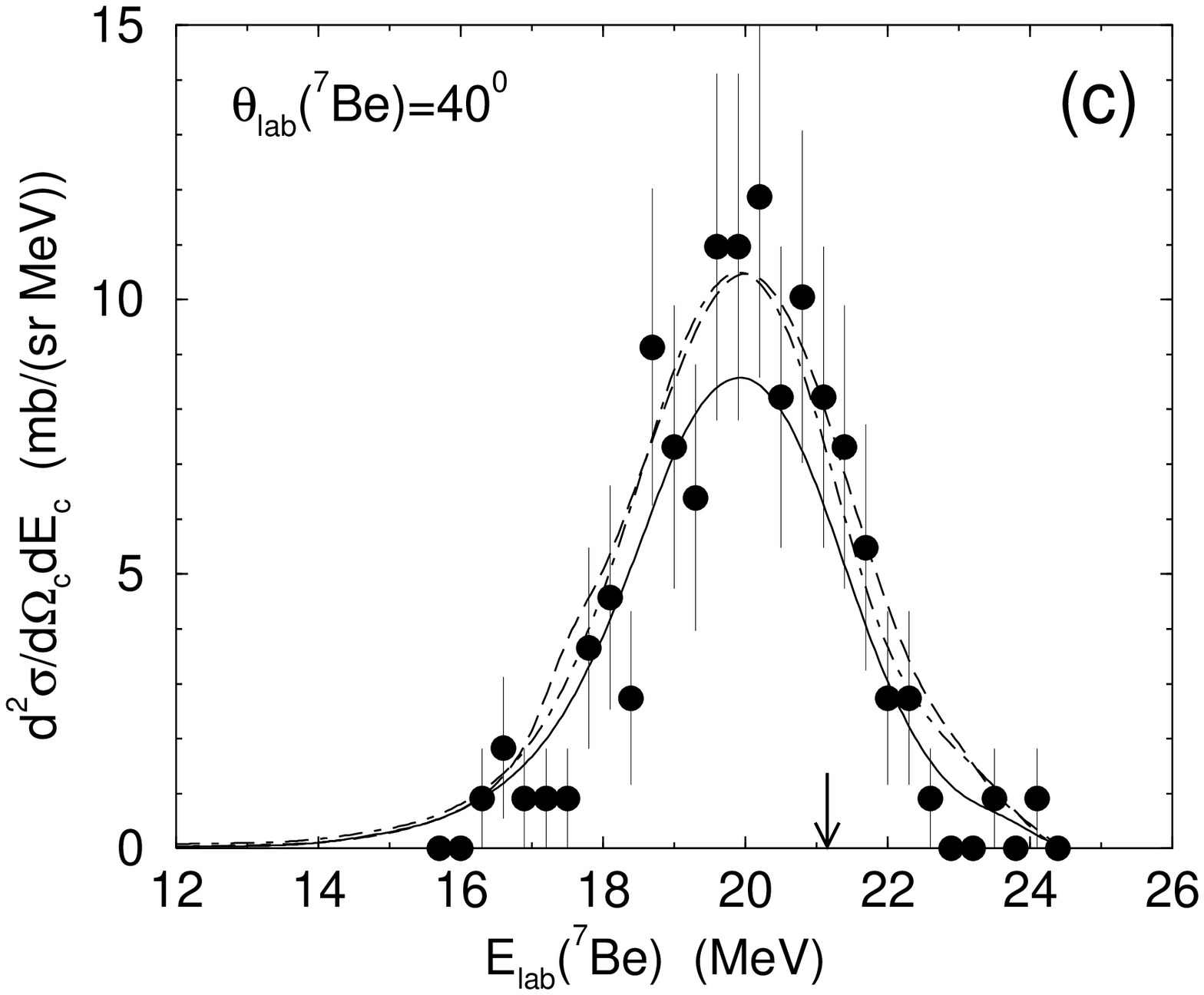,width=0.45\textwidth}}
\hspace{0.4in}
        \parbox[t]{0.4\textwidth}{
\epsfig{figure=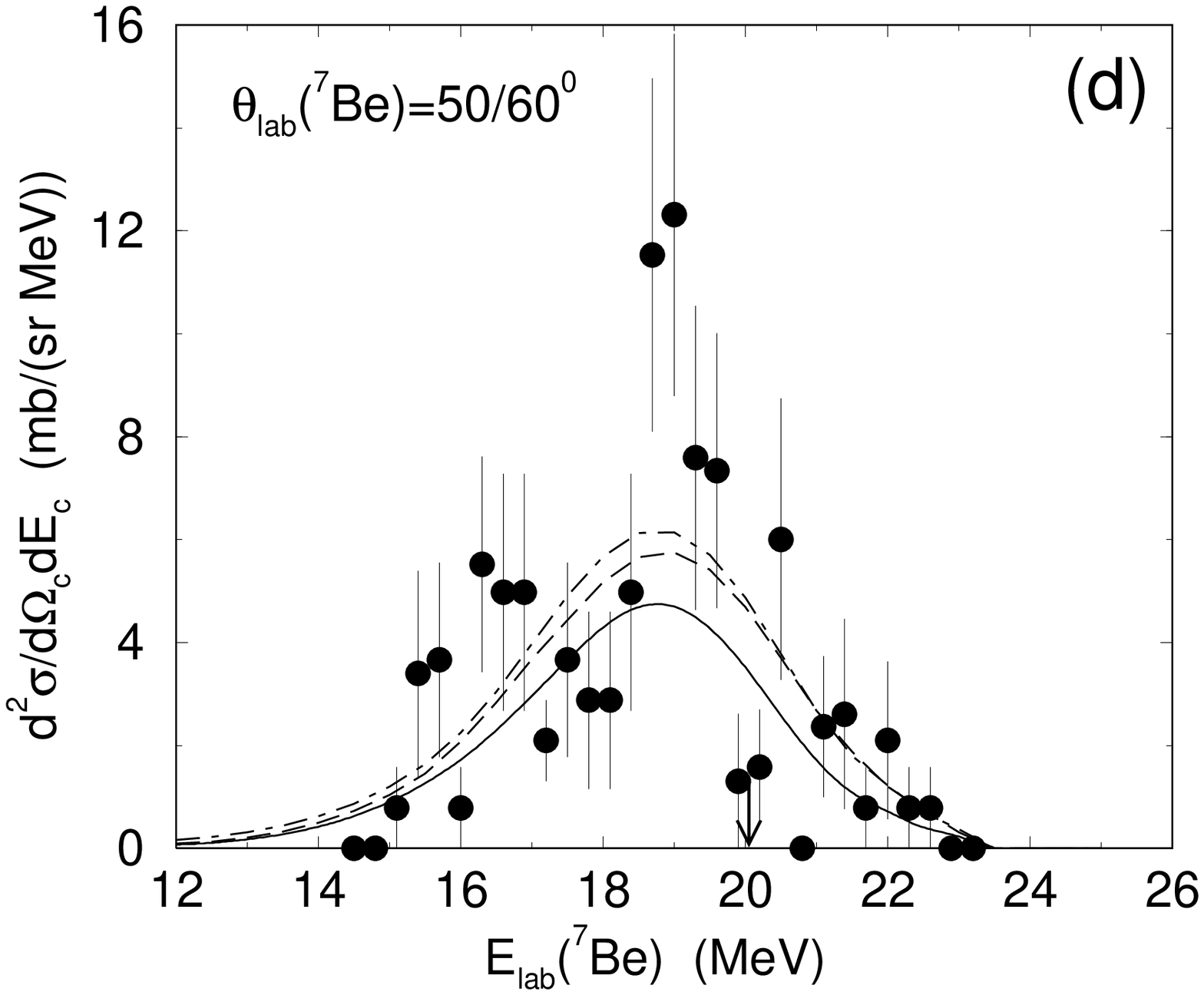,width=0.45\textwidth}}
}
\medskip
\caption{Calculated laboratory frame $^7$Be cross section energy distributions
following the breakup of $^8$B on $^{58}$Ni at 25.8 MeV for the laboratory
angles indicated. The calculations use the EB (solid) and Kim (dashed) models
for the  proton-$^7$Be interaction and the BG proton-target interaction. The
dot-dashed curves use the EB proton-$^7$Be interaction and the VG proton-target
interaction. The arrows on the energy axis indicate 7/8 of the $^8$B energy for
elastic scattering at each laboratory angle. The experimental data are from
Ref.\ \protect\cite{nd2}.} \label{fig:exp}
\end{figure}

\begin{figure}[h]
\centerline{
\hspace{-0.5in}
        \parbox[t]{0.4\textwidth}{
\epsfig{figure=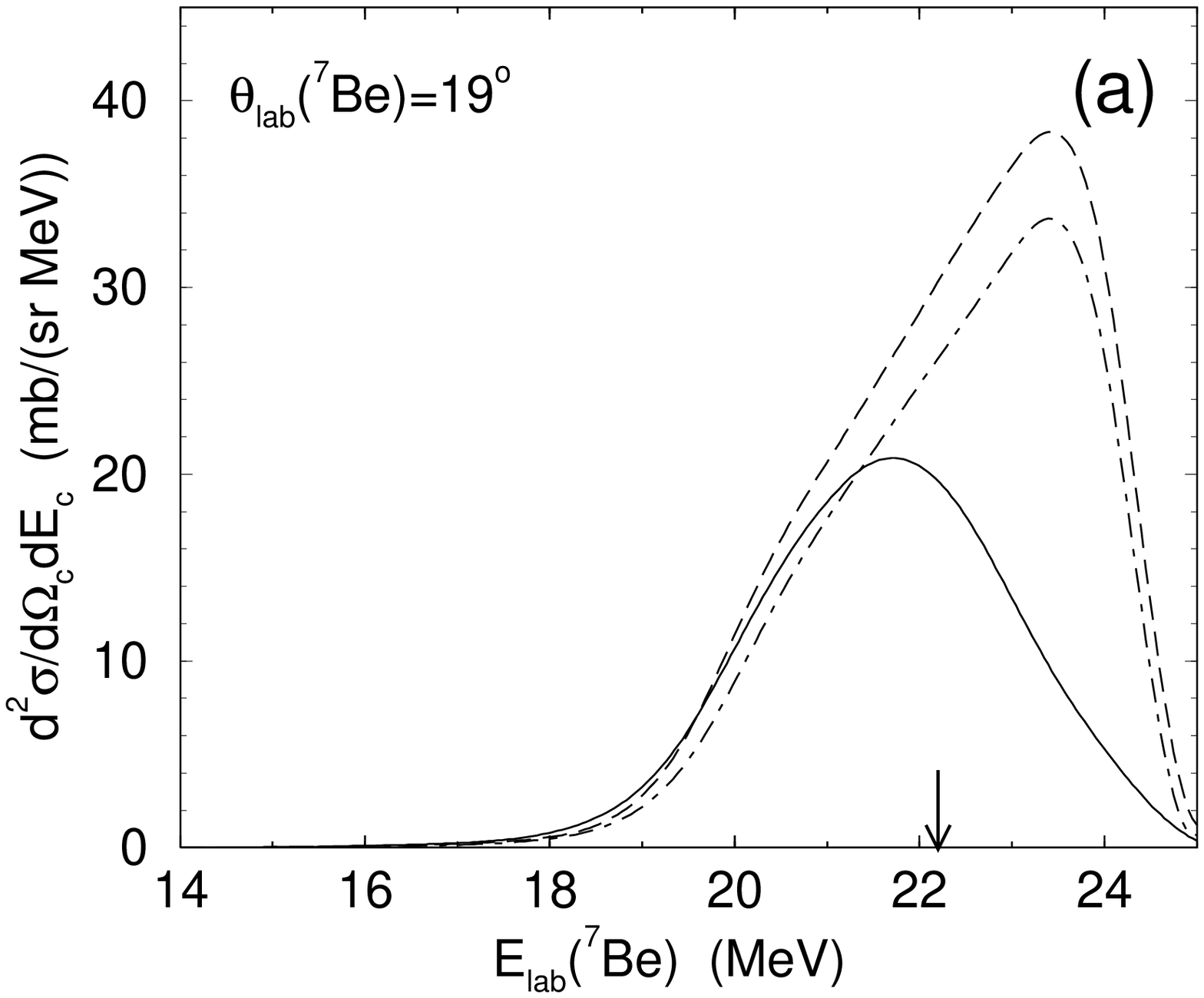,width=0.45\textwidth}}
\hspace{0.4in}
        \parbox[t]{0.4\textwidth}{
\epsfig{figure=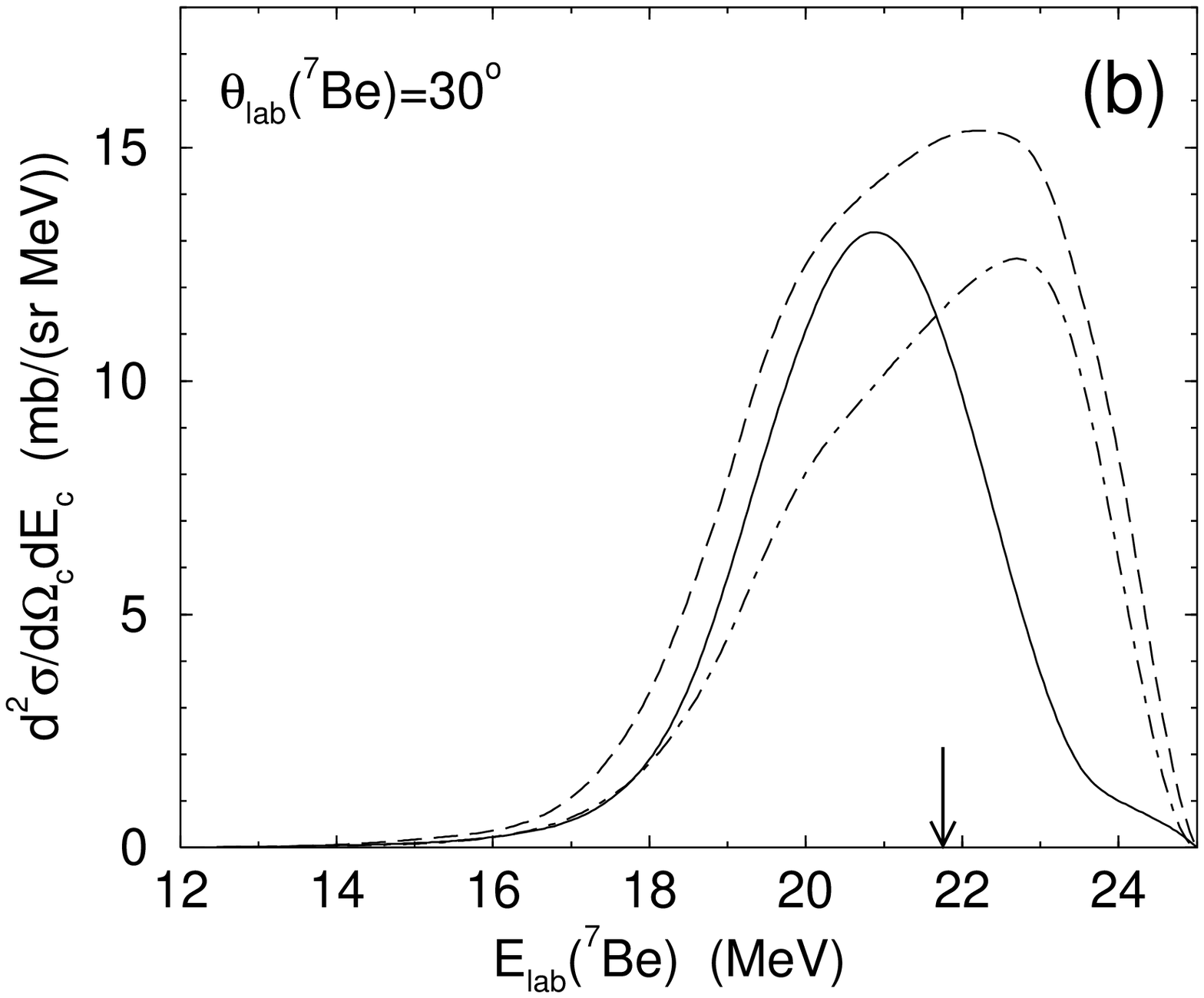,width=0.45\textwidth}}
}
\centerline{
\hspace{-0.5in}
        \parbox[t]{0.4\textwidth}{
\epsfig{figure=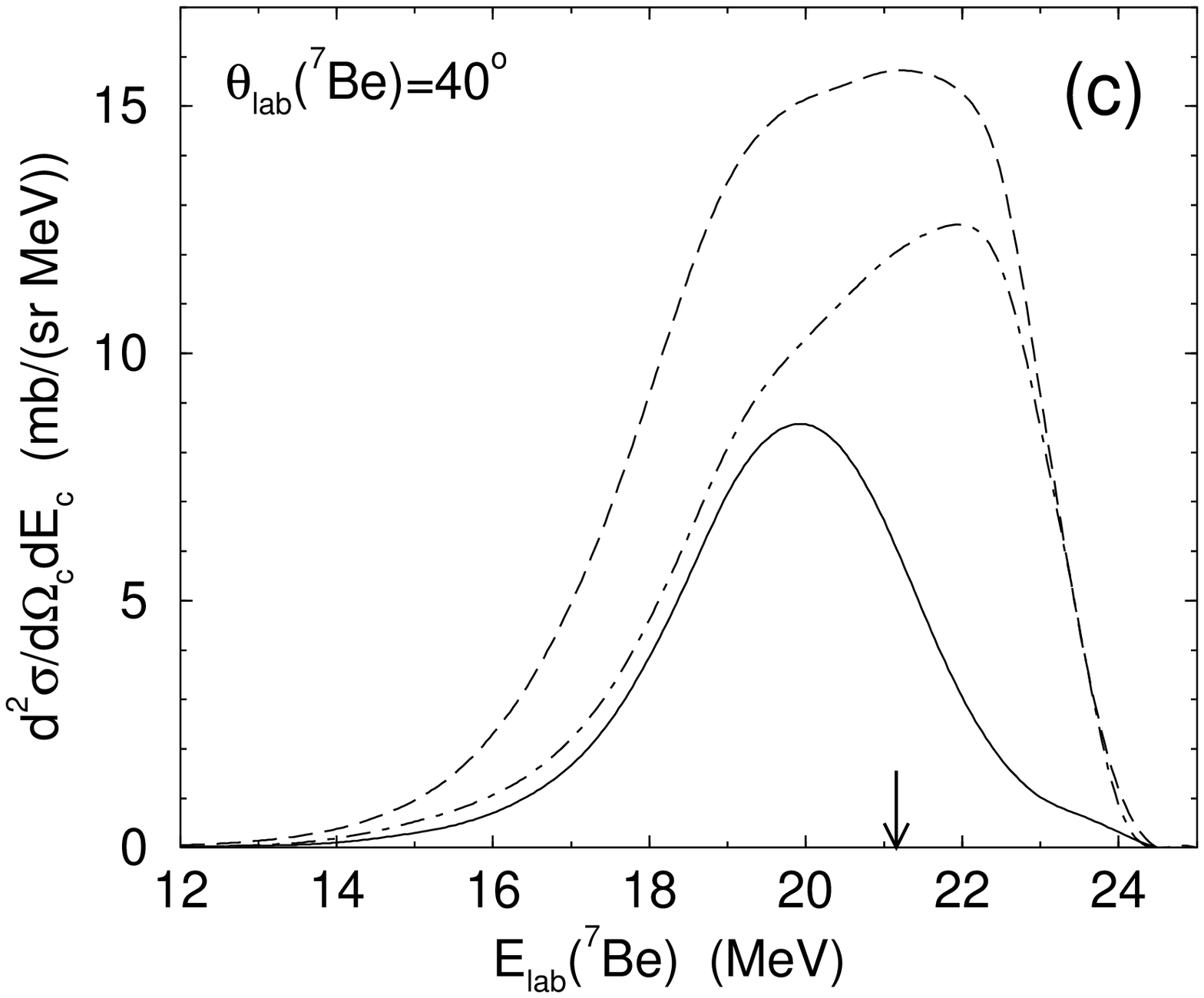,width=0.45\textwidth}}
\hspace{0.4in}
        \parbox[t]{0.4\textwidth}{
\epsfig{figure=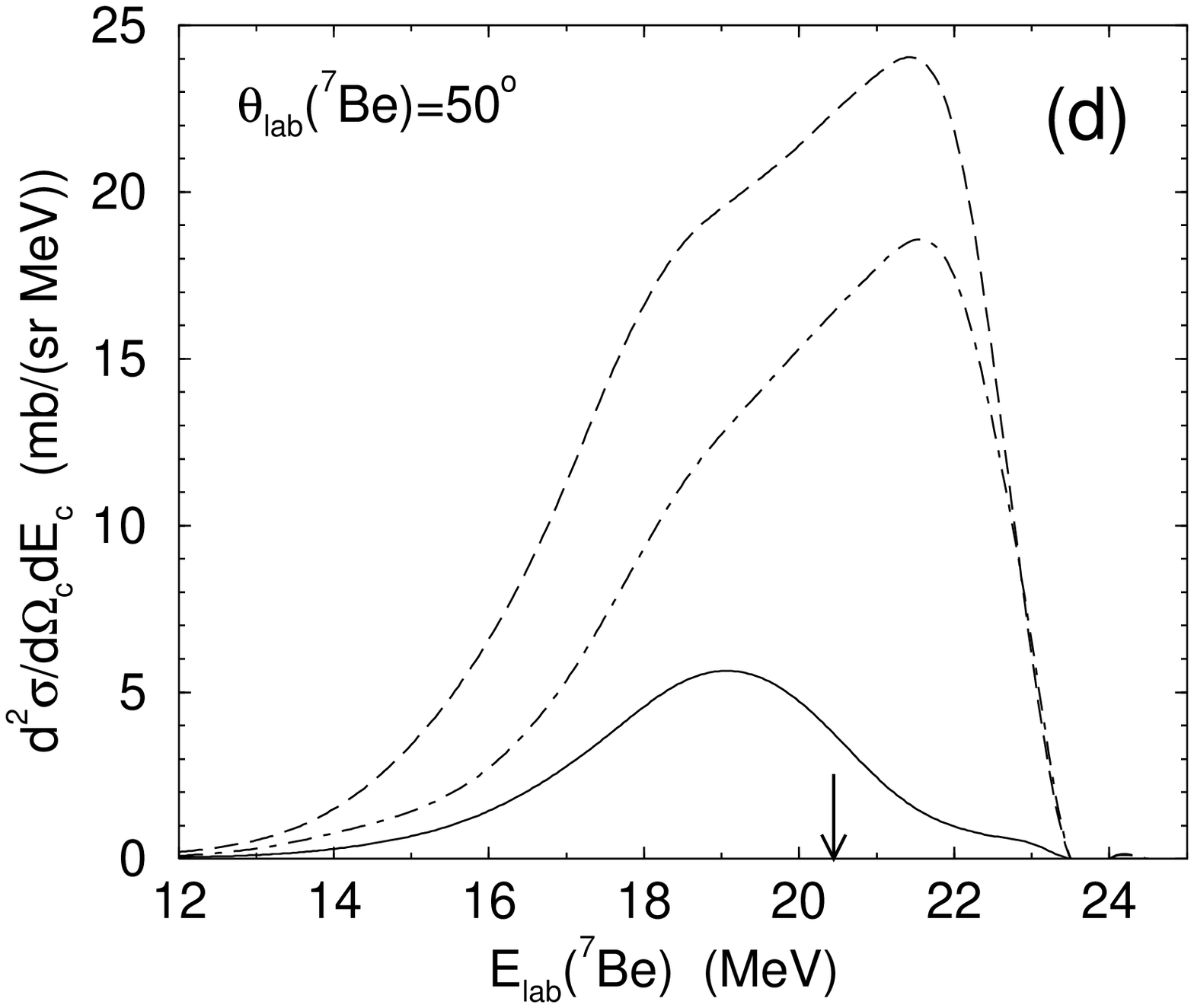,width=0.45\textwidth}}
}
\medskip
\caption{Calculated laboratory frame $^7$Be cross section energy distributions
following the breakup of $^8$B on $^{58}$Ni at 25.8 MeV for the laboratory
angles indicated. The curves compare the full CDCC (solid), the CDCC in the
absence of the CC bin couplings (dot-dashed), and the DWBA (long-dashed)
calculations. All calculations use the EB $^8$B ground state structure model
and the BG proton distortion. The arrows on the energy axis indicate 7/8 of the
$^8$B energy for elastic scattering at each laboratory angle.} \label{fig:th}
\end{figure}

In Fig.\ \ref{fig:exp} we show the calculated breakup energy distributions of
the $^7$Be fragments, together with the data from Ref.\ \cite{nd2}, for four
measured laboratory configurations.  For the smallest angle, $\approx
20^\circ$, the calculations and the data are the average of the distributions
at $\theta_{lab}=19^\circ$ and $\theta_{lab}= 21^\circ$.  For the largest
angles, $50/60^\circ$, the curves and data are similarly the average of the
distributions obtained at $\theta_{lab}=50^\circ$ and $\theta_{lab}=
60^\circ$.  The measured cross sections are zero outside of the range of the
data points shown. The solid curves use the BG proton distortion and the EB
proton-$^7$Be potential. The general features of the data, their magnitude,
centroids, and widths, are well described by the calculations. The long-dashed
curves are the results using the Kim proton-$^7$Be potential. They show an
enhanced cross section, discussed earlier, but a very similar shape. The
dot-dashed curves are calculated using the VG proton distortion and the EB
proton-$^7$Be potential. The small arrows on the energy axis in
Fig.\ \ref{fig:exp} (and Fig.\ \ref{fig:th}) indicate 7/8 of the $^8$B energy
for elastic scattering at each laboratory angle. An overall reduction in the
mean energy of the heavy fragments within the breakup reaction is evident.

\begin{figure}[h]
\centerline{
\hspace{-0.5in}
        \parbox[t]{0.4\textwidth}{
\epsfig{figure=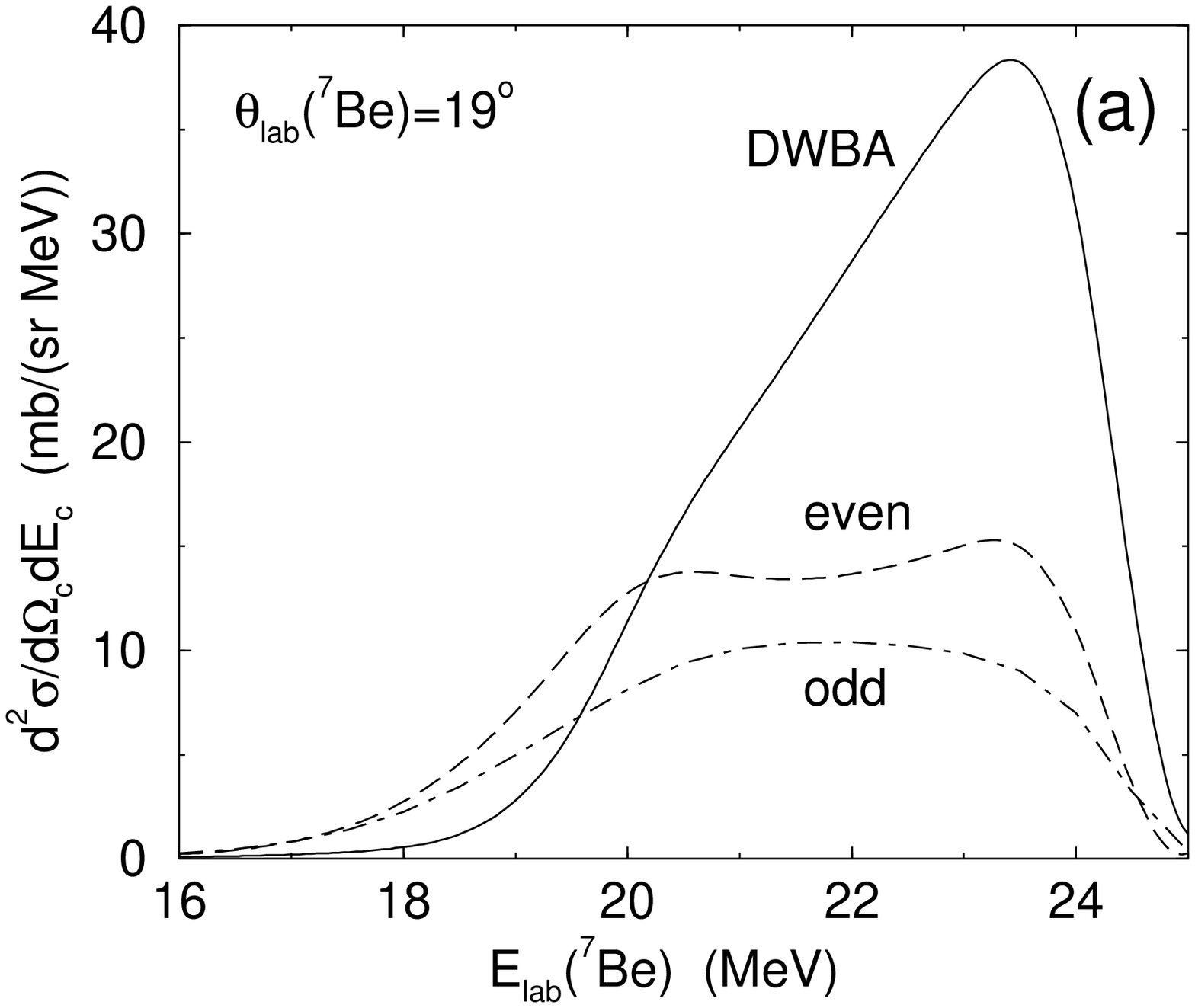,width=0.45\textwidth}}
\hspace{0.4in}
        \parbox[t]{0.4\textwidth}{
\epsfig{figure=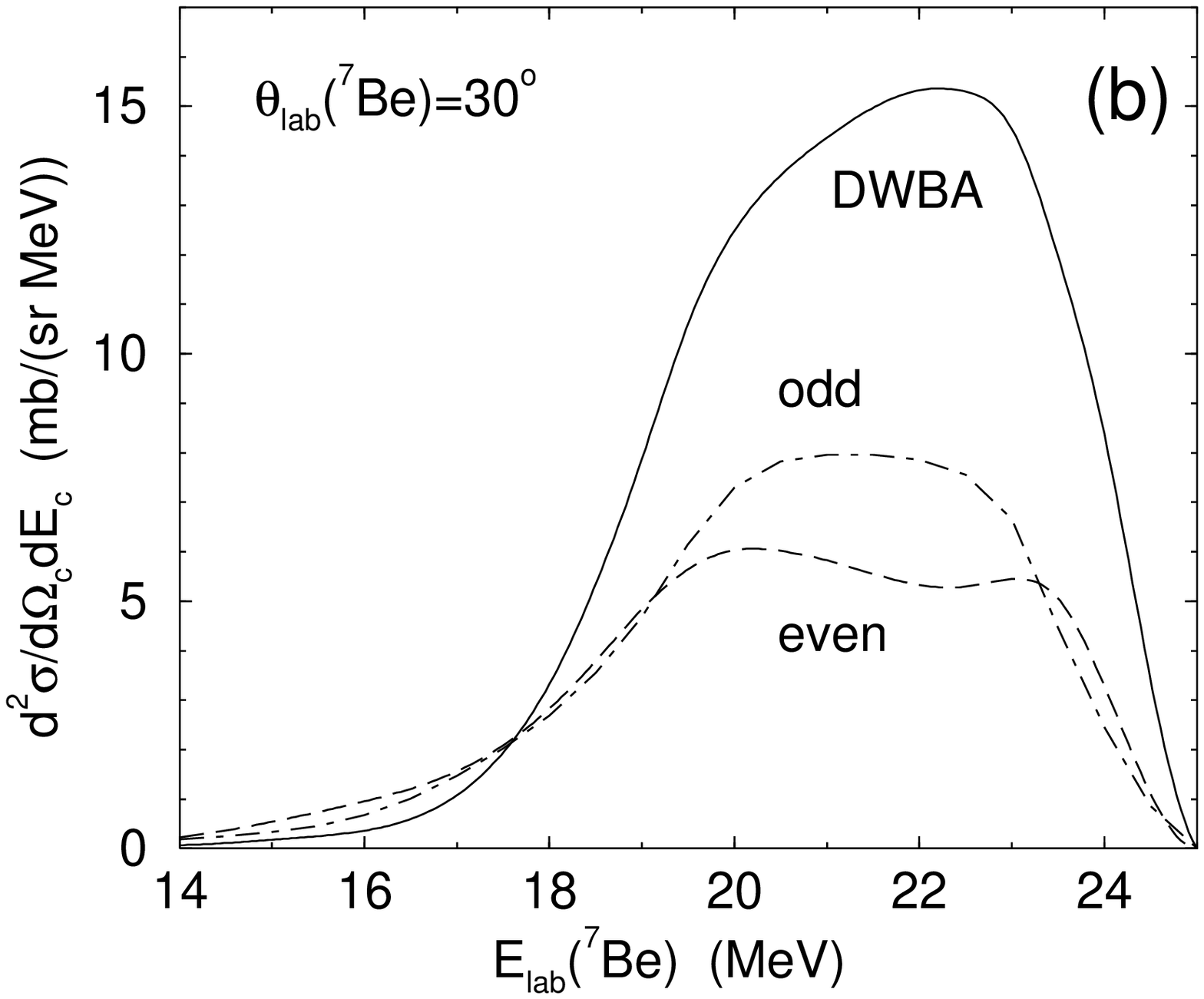,width=0.45\textwidth}}
}
\medskip
\caption{Calculated laboratory frame $^7$Be cross section energy distributions
following the breakup of $^8$B on $^{58}$Ni at 25.8 MeV for the laboratory
angles indicated. The curves show the separate odd and even breakup partial
waves cross sections and their interference within the full DWBA calculation.}
\label{fig:e1e2}
\end{figure}

Further insight is gained by looking at the results of DWBA calculations, and
also calculations in which a subset of the continuum couplings are switched
off, shown in Figs.\ \ref{fig:th}(a--d). The long-dashed lines show the DWBA
calculations. The dot-dashed lines are the results of CDCC coupled channels
calculations but in which all continuum-continuum (CC) couplings between bin
states are removed. The solid lines are the full calculations, as were shown in
Fig.\ \ref{fig:exp}. We see that the calculations in the absence of CC
couplings, both DWBA and truncated coupled channels, show energy distributions
which are strongly asymmetric and have an enhanced high energy peak. This is
very similar to what is observed in the $^7$Be fragment parallel momentum
distributions from $^8$B breakup observed at higher-energy \cite{davids}. As in
that case, we show in Fig. \ref{fig:e1e2} that this asymmetry has its origin in
the interference between the E1 transitions to even breakup partial waves, and
the E2 transitions to odd breakup partial waves. These $E\lambda$ amplitudes,
which individually give approximately symmetric energy distributions, interfere
to give strongly asymmetric responses. The very nearby kinematic cutoff in our
case distorts the symmetry somewhat.  The E2/E1 amplitude ratio in this lower
energy case is also greater and so the asymmetry is enhanced compared to higher
energies.

\begin{figure}[h]
\epsfig{figure=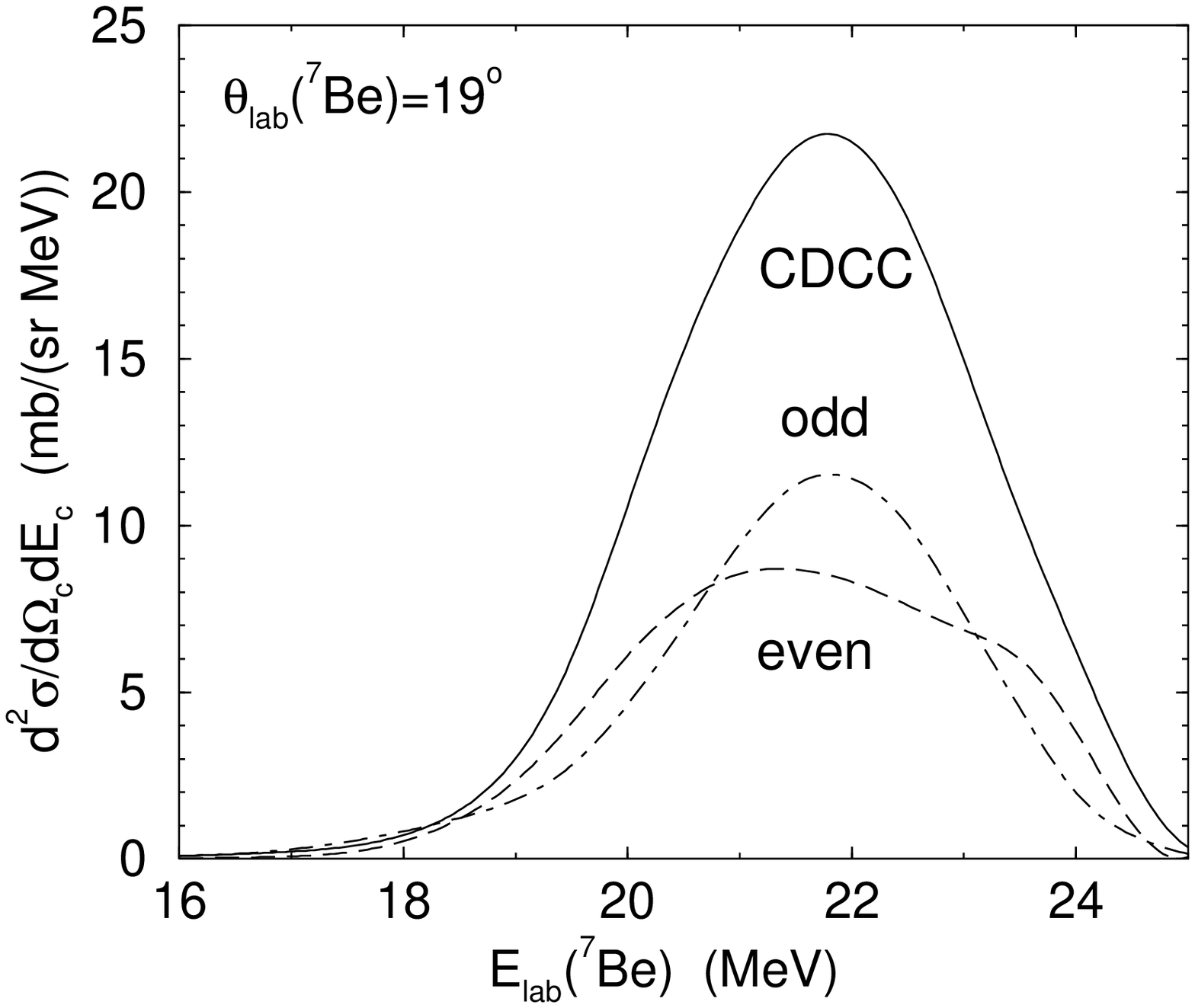,width=0.60\textwidth}
\medskip
\caption{Calculated laboratory frame $^7$Be cross section energy distributions
following the breakup of $^8$B on $^{58}$Ni at 25.8 MeV for the laboratory
angle indicated. The curves show the odd and even breakup partial waves cross
sections and their interference within the full CDCC calculation.}
\label{fig:ec1b}
\end{figure}

In the full CDCC calculations these asymmetries are essentially removed as a
result of the higher-order couplings. This higher-order coupling induced
suppression of the E1/E2 interference asymmetry was also a feature of the
(higher energy) dynamical calculations in Ref.\ \cite{esb}. The suppression is
more complete at the lower energy discussed here. Figure \ref{fig:ec1b} shows
the analogue of Fig.\ \ref{fig:e1e2}(a), the calculated cross sections to odd
and even breakup partial waves, from the full CDCC calculations using EB and BG
potentials. Evident is the interference, both within and between the odd and
even partial wave excitations. We note that the analogue of the $E2$ cross
section, the $p+f$ wave contribution, is not itself suppressed, and is in fact
large. The interference between the two contributions in Figs.\  \ref{fig:ec1b}
and \ref{fig:e1e2}(a) is however very different in the two cases.

Also evident in these two figures is that the odd breakup partial waves
contribution in the CDCC calculation is significantly narrower than that
calculated using DWBA. This narrowing is already manifest in $s+p$ wave
two-step ($q\leq2$ Coulomb) calculations and arises there from interference
between the first-order E2 and second-order E1 amplitudes for populating the
$p$ wave continuum. The importance of these particular interfering paths was
also noted in Ref.\ \cite{esb}, there in connection with a reduction in the
calculated $^8$B decay energy spectrum at higher energy, when going beyond
first-order Coulomb excitation theory. The calculated energy distributions
reveal even more clearly than those for the angular distribution the importance
of a full treatment of the dynamical couplings within the continuum.

\subsubsection{Additional calculations and comments}

Since the proton separation energy from the $^{59}$Cu(g.s.) is $S_p = 3.42$
MeV, proton transfer to the $^{59}$Cu(g.s.) would produce $^7$Be fragments with
$\approx 26$ MeV of kinetic energy in the c.m.\ frame, and so such events are
not part of the energy distributions measured.  Those transfers that might
contribute to the energy spectra of Fig. \ref{fig:exp} would therefore be to
excited (resonant) proton levels in $^{59}$Cu$^*$ at around 9 MeV of
excitation. If the proton-$^{58}$Ni interaction supported one or more potential
resonances, then the CDCC reaction mechanism would include their dynamical
effects since breakup, by projectile excitation and by transfer to unbound
states, are not distinguishable mechanisms in the three-body reaction model
used.  Clearly, however, the ability of the proton-$^{58}$Ni interaction to
support such resonance strength, and its absorptive content, are closely
related questions.  As was noted earlier, Fig.  \ref{fig:ang}, use of the VG
proton-target potential calculates an enhanced large angle cross section.
Clarifying this sensitivity, and the possible role of such final-state
resonances, requires further study and fine tuning of the proton-target
potential. A full discussion of this topic is beyond the scope and motivation
of the present article.

\begin{figure}[h]
\epsfig{figure=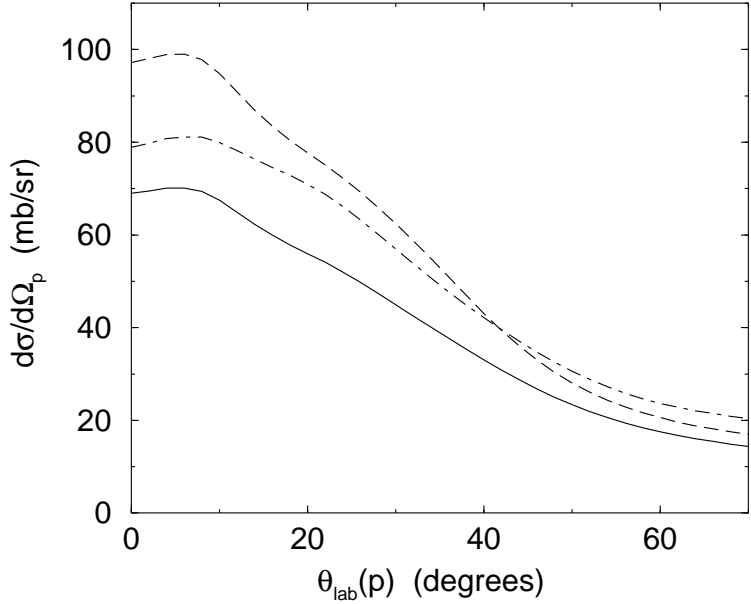,width=0.60\textwidth}
\medskip
\caption{Calculated laboratory frame proton cross section angular distributions
following the breakup of $^8$B on $^{58}$Ni at 25.8 MeV, showing the role of
the interaction between the proton and the target. The calculations use the EB
(solid) and Kim (long dashed) models for the  proton-$^7$Be interaction and the
BG proton-target interaction. The dot-dashed curve uses the EB proton-$^7$Be
interaction and the VG proton-target interaction.} \label{fig:pang}
\end{figure}

With this sensitivity to the proton-target potential in mind, however, in
Fig.\ \ref{fig:pang}, we show the calculated proton laboratory angular
distributions from the EB and Kim $^8$B wave functions, and the BG and VG
proton distorting potentials. We note that the magnitude, but not the shape, of
the proton cross section angular distribution shows a significant sensitivity
to the assumed absorption in the proton-target system. Precise data could
therefore verify and constrain this element of the calculations.

\begin{figure}[h]
\epsfig{figure=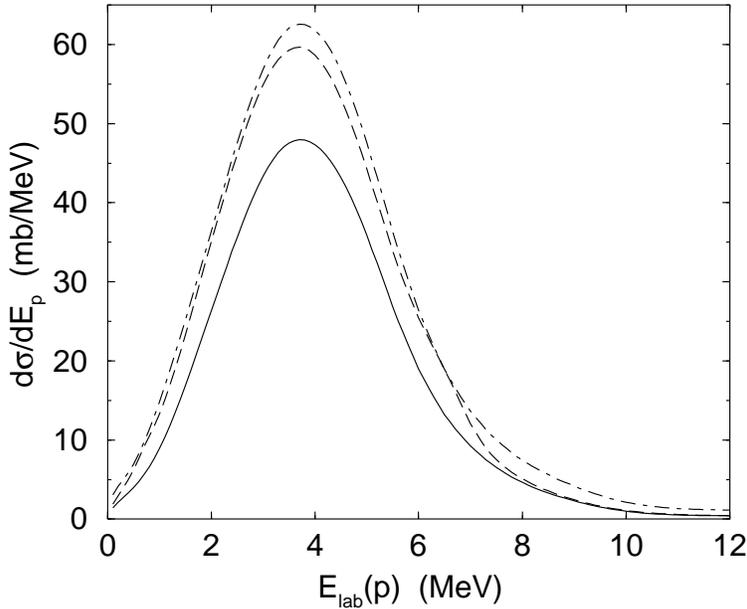,width=0.60\textwidth}
\medskip
\caption{Calculated laboratory frame angle integrated proton cross section
energy distributions following the breakup of $^8$B on $^{58}$Ni at 25.8 MeV.
The calculations use the EB (solid) and Kim (long dashed) models for the
proton-$^7$Be interaction and the BG proton-target interaction. The dot-dashed
curve uses the EB proton-$^7$Be interaction and the VG proton-target
interaction.} \label{fig:prot.en}
\end{figure}

\begin{figure}[h]
\epsfig{figure=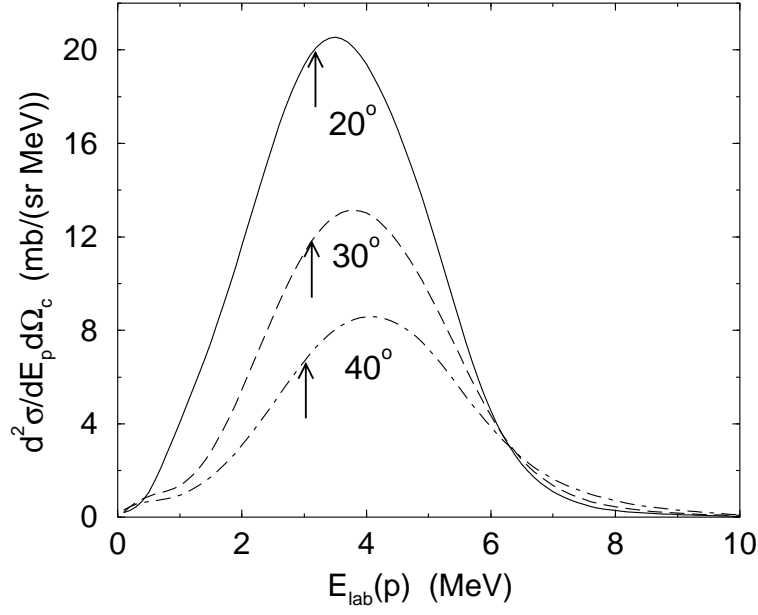,width=0.60\textwidth}
\medskip
\caption{Calculated laboratory frame proton cross section energy distributions
following the breakup of $^8$B on $^{58}$Ni at 25.8 MeV for the $^7$Be fragment
laboratory angles indicated. The calculations use the EB proton-$^7$Be
interaction and the BG proton-target interaction. The arrows indicate 1/8 of
the $^8$B energy for elastic scattering at each laboratory angle.}
\label{fig:prot.en.7}
\end{figure}

The shape of the calculated proton energy distribution, like that for the
$^7$Be fragments, shows little sensitivity to the absorptive content of the
proton distortion or to the choice of $^8$B binding potential. The calculations
in Fig.\ \ref{fig:prot.en} use the EB (solid) and Kim (long dashed) models for
the proton-$^7$Be interaction and the BG proton-target interaction. The
dot-dashed curve uses the EB proton-$^7$Be interaction and the VG proton-target
interaction. The calculated proton energy distributions, integrated over all
$^7$Be fragment angles, peak for $E_p \approx 3.8$ MeV and have a width $\Gamma
\approx 4$ MeV. The tail of the energy distribution is seen to extend to high
energy, reflecting the high relative energy excitations of the $^8$B$^*$
discussed earlier in connection with the convergence of the CDCC calculations.
Figure\ \ref{fig:prot.en.7} shows the energy distributions predicted when the
$^7$Be fragments emerge at laboratory angles of 20, 30 and 40$^\circ$. In this
case the arrows on the different curves indicate 1/8 of the $^8$B energy for
elastic scattering at each laboratory angle. The calculations show an increased
average energy (acceleration) of the removed protons from the dynamics of the
breakup process.

\section{Summary and conclusions}

In this paper we have calculated the most exclusive three-body breakup
observables of a two-body projectile using the coupled channels CDCC
methodology.  The formalism is applied to investigate the angular and energy
distributions of the $^7$Be fragments resulting from the sub-Coulomb breakup of
$^8$B on a $^{58}$Ni target, the subject of recent experiments. We show that
the convergence of the CDCC calculations of these observables is more subtle
than that for the cross section of the c.m.\ motion of the $^8$B$^*$ and
requires a significantly more extended space of $^8$B$^*$ excitation energies.
The required excitation energy range is clarified.

Our calculations show that the $^8$B structure and the absorptive content of
the proton-target potentials affect the calculated $^7$Be fragment angular
distributions differently, the former producing an overall scaling, and the
latter a large angle enhancement. Reducing the strength of the imaginary part
of the proton potential in line with a phenomenological study \cite{van61},
provides agreement with the larger angle data. The full CDCC calculations are
shown to provide a good description of the measured $^7$Be fragment energy
distributions. The widths and positions of these distributions are found to be
rather insensitive to the details of the potentials used within the
calculations. The presence of coupling between the continuum states is shown to
be crucial to understand both the magnitudes of these energy distributions and
their measured energy centroids.  The absorptive content of the proton-target
potentials affect the magnitudes of the calculated proton angular and energy
distributions significantly, although their shapes are little affected. The
calculated proton ($^7$Be) fragment energy distribution reveals an overall
increased (reduced) average energy of the fragment from the dynamics of the
breakup process.

The application of these techniques to calculate the parallel momentum
distribution of the heavy breakup fragments following the nuclear dissociation
of the two-body system $^{11}$Be will be reported elsewhere \cite{jat00}.
Further applications to systems with significant Coulomb dissociation strength,
such as for $^8$B breakup at energies of 40 MeV/nucleon and greater, are also
in progress.

\acknowledgements
We thank Dr Valdir Guimar\~aes and Prof. Jim Kolata for providing the data
presented in tabular form and for detailed discussions of the experimental
arrangement. The financial support of the United Kingdom Engineering and
Physical Sciences Research Council (EPSRC) in the form of Grants
Nos.\ GR/J95867 and GR/M82141 and Portuguese support from Grant FCT
PRAXIS/PCEX/P/FIS/4/96 are gratefully acknowledged.

\newpage

\newpage

\newpage

\newpage

\newpage

\end{document}